\documentclass[twocolumn,showpacs,preprintnumbers,%
               aps,prl]{revtex4-1}
\usepackage[dvips]{graphicx}
\usepackage{amsmath,amssymb}
\usepackage[breaklinks, colorlinks=true, pdfstartview=FitV, linkcolor=red,%
 citecolor=blue, urlcolor=blue]{hyperref}

\newcommand{\tr}{\text{tr}}
\newcommand{\Nc}{N_{\text{c}}}
\newcommand{\Nf}{N_{\text{f}}}

\newcommand{\Lag}{\mathcal{L}}
\newcommand{\calA}{\mathcal{A}}
\newcommand{\calM}{\mathcal{M}}
\newcommand{\calP}{ $\mathcal{P}$ }
\newcommand{\calCP}{ $\mathcal{CP}$ }
\newcommand{\LQCD}{\Lambda_{\text{QCD}}}

\newcommand{\bq}{\boldsymbol{q}}

\newcommand{\bB}{\boldsymbol{B}}

\begin{document}

\title{Wess-Zumino-Witten action and photons from
       the Chiral Magnetic Effect}

\author{Kenji Fukushima}
\affiliation{Department of Physics, Keio University,
             Kanagawa 223-8522, Japan}
\author{Kazuya Mameda}
\affiliation{Department of Physics, Keio University,
             Kanagawa 223-8522, Japan}

\begin{abstract}
 We revisit the Chiral Magnetic Effect (CME) using the chiral
 Lagrangian.  We demonstrate that the electric-current formula of the
 CME is derived immediately from the contact part of the
 Wess-Zumino-Witten action.  This implies that the CME could be, if
 observed, a signature for the local parity violation, but a direct
 evidence for neither quark deconfinement nor chiral restoration.  We
 also discuss the reverse Chiral Magnetic Primakoff Effect, i.e.\ the
 real photon production through the vertex associated with the CME,
 which is kinematically possible for space-time inhomogeneous
 configurations of magnetic fields and the strong $\theta$ angle.  We
 make a qualitative estimate for the photon yield to find that it
 comparable to the thermal photon.
\end{abstract}
\maketitle


The vacuum structure in Quantum Chromodynamics (QCD) has been an
important subject investigated in theory for a long time.  It has been
well-known that gauge configurations with topologically non-trivial
winding such as the instanton, the magnetic monopole, etc should play
a crucial role in the spontaneous breaking of chiral
symmetry~\cite{Schafer:1996wv}, color
confinement~\cite{Diakonov:2009jq}, the mass of $\eta'$
meson~\cite{'tHooft:1986nc}, and the strong $\theta$
angle~\cite{Kim:2008hd}.

Among others the problem of the strong $\theta$ angle is still posing
a theoretical challenge.  There is no consensus on the unnatural
smallness of $\theta$ and thus the absence of \calP and \calCP
violation in the strong interaction.  Recently, more and more
researchers in the field of the relativistic heavy-ion collision are
getting interested in the possibility of fluctuating $\theta$ in a
transient state of QCD matter and searching for a signature to detect
the local \calP violation (LPV)
experimentally~\cite{Morley:1983wr,*Kharzeev:1998kz,%
*Fugleberg:1998kk,*Buckley:1999mv,Voloshin:2000xf,*Voloshin:2004vk,%
*Finch:2001hs,*Abelev:2009ac}.

In this context the discovery of the Chiral Magnetic Effect
(CME)~\cite{Kharzeev:2007tn,*Kharzeev:2007jp,Fukushima:2008xe} has
triggered constructive discussions and lots of works have been devoted
to the interplay between the topological effects and the external
magnetic field
$\bB$~\cite{Son:2007ny,Mizher:2008hf,*Mizher:2010zb,Asakawa:2010bu},
while the strong-$\bB$ effect itself on nuclear or QCD matter had
been~\cite{Rafelski:1975rf} and are
still~\cite{D'Elia:2010nq,*Bali:2011qj} attracting theoretical
interest.
(See Ref.~\cite{Vilenkin:1980fu,*Metlitski:2005pr} for earlier works
related to the CME.)  If $\theta$ temporarily takes a non-zero value
in hot and dense QCD matter, its time derivative induces an excess of
either left-handed or right-handed quarks.  Because of the alignment
of the spin and the momentum directions of left-handed and
right-handed quarks, $\bB$ would generate a net electric current
parallel to $\bB$, which may be in principle probed by the
fluctuations of $\mathcal{P}$-odd observables in the heavy-ion
collision~\cite{Voloshin:2000xf,*Voloshin:2004vk,%
*Finch:2001hs,*Abelev:2009ac}.

It should be an urgent problem of paramount importance, we believe, to
sort out the proper physics interpretation of the CME and the LPV in
general since the LPV is under intensive investigations in ongoing
experiments at present.  It is also under active discussions whether
the Chiral Magnetic Wave (CMW) should account for the discrepancy
between the elliptic flows of positively and negatively charged
hadrons~\cite{Kharzeev:2010gd,*Burnier:2011bf}.  It is often said that
the CME could be a signature for quark deconfinement and chiral
symmetry restoration, as stated also by one of the present authors in
Ref.~\cite{Fukushima:2008xe}.  This was conjectured because the
intuitive explanation for the CME seemed to require almost massless
$u$ and $d$ quarks.  One should be, however, careful of the physics
interpretation of anomalous phenomena which sometimes look
counter-intuitive.  The first half of our discussions is devoted to
considerations on the implication of the CME in terms of the chiral
Lagrangian.  We conclude that the CME is insensitive to whether the
fundamental degrees of freedom are quarks or hadrons, so that it could
be seen without deconfinement.  Chiral symmetry restoration is, on the
other hand, necessary to realize the hadronic LPV in the same manner
as in the case of the disoriented chiral condensate
(DCC)~\cite{Anselm:1989pk,*Anselm:1991pi,*Blaizot:1992at,*Rajagopal:1993ah}.

In the last half of our discussions, as an application of the chiral
Lagrangian, we address the real photon production through the process
that we call the reverse Chiral Magnetic Primakoff Effect.  The
typical process in the ordinary Primakoff effect is the $\pi^0$ (or
some neutral meson generally) production from a single photon picking
up another photon from the external electromagnetic
field~\cite{Primakoff:1951pj}.  In the relativistic heavy-ion
collision the neutral pseudo-scalar field $\theta(x)$ can couple to a
photon in $\bB$ leading to a single photon emission,
i.e.\ $\theta + B \to \gamma$, which can be viewed as a reverse
process of the Primakoff effect.  Such a mechanism for the photon
production can be traced back to the old idea to detect the axion via
the Primakoff effect~\cite{Sikivie:1983ip,*Raffelt:1987im}, and is
similar to the recent idea on a novel source of photons from the
conformal anomaly~\cite{Basar:2012bp}.  In short, a crucial difference
between our idea and that in Ref.~\cite{Basar:2012bp} lies in the
neutral meson involved in the process --- $\sigma$ meson (which turns
to a hydrodynamic mode) in the conformal anomaly case and $\theta$ or
$\eta_0$ in our case of the CME vertex (see also
Ref.~\cite{Fukushima:2002mp} for the diphoton emission from the
$\sigma$ meson).  From this point of view, it would be very natural to
think of photons as a signature of the CME;  instead of the
axion~\cite{Sikivie:1983ip}, CME requires a \textit{background}
$\theta(x)$, which may cause the same process of the single photon
production as the axion detection.

The interesting point in our arguments for the photon production is
that the real photon emission is attributed to exactly the same vertex
as to describe the electric-current generation in the CME.\ \ As long
as $\bB$ and $\theta$ are spatially homogeneous, as often assumed for
simplicity, the real particle production is prohibited kinematically,
but once $\bB$ and $\theta$ are space-time dependent (and they are
indeed so in the heavy-ion collision!), the energy-momentum
conservation is satisfied, so that the real photon can come out.


Here, one might have wondered how the physical constant $\theta$ can
be lifted up in hot and dense matter and treated as if it were a
particle.  In other words, what is the origin of the chiral chemical
potential $\mu_5$ in the hadronic environment?  This is an important
question and related to the physical mechanism to cause the LPV.\ \ At
extremely high energy the Color Glass Condensate and the Glasma
initial condition~\cite{Kharzeev:2001ev} may be the most relevant and
its characteristic scale is then given by the saturation scale $Q_s$.
In this case the role of $\theta$ in the pure Yang-Mills dynamics is
more non-trivial~\cite{D'Elia:2012vv} than full QCD with dynamical
quarks where $\theta$ can be regarded as the $\mathrm{U(1)_A}$
rotation angle.  In the hadronic phase at low energy, the chiral
Lagrangian provides us with a clear picture, which consists of three
parts,
\begin{equation}
 \Lag_{\text{eff}} = \Lag_\chi + \Lag_{\text{WZW}} + \Lag_{\text{P}}\;,
\end{equation}
where the first one is the usual chiral Lagrangian that is given
by~\cite{DiVecchia:1980ve,*Witten:1980sp,*Leutwyler:1992yt,%
Kaiser:2000gs,*Kaiser:2000ck}
\begin{equation}
 \begin{split}
 \Lag_\chi &= \frac{f_\pi^2}{4} \tr\bigl[ D_\mu U^\dagger
  D^\mu U + 2\chi (MU^\dagger + UM ) \bigr] \\
 &\qquad\qquad -\frac{\Nf \chi_{\text{top}}}{2}
  \Bigl[\theta-\frac{i}{2}\tr(\ln U - \ln U^\dagger)
  \Bigr]^2\;,
 \end{split}
\end{equation}
in the lowest order including the topological terms that break
$\mathrm{U(1)_A}$ symmetry.  Here, $\chi_{\text{top}}$ represents the
pure topological susceptibility, the covariant derivative involves the
vector and the axial-vector fields as
$D_\mu U \equiv \partial_\mu U - i r_\mu U + i U l_\mu +
 \frac{i}{2}(\partial_\mu \theta + 2\tr(a_\mu)) U$ with
$r_\mu \equiv v_\mu + a_\mu$ and $l_\mu = v_\mu - a_\mu$, and
$\chi\equiv -\langle\bar{q}q\rangle/f_\pi^2$ from the
Gell-Mann-Oakes-Renner relation.  It is obvious that, as discussed in
Ref.~\cite{DiVecchia:1980ve,*Witten:1980sp,*Leutwyler:1992yt}, the
$\theta$-dependence is to be absorbed in the phase of $U$ if the
current quark mass matrix $M$ has a zero component.  Then, one can
understand that $\theta$ and the phase of $U$ or $\eta_0/f_{\eta_0}$
are simply identifiable apart from the mass terms proportional to
$\chi$ and $M$.  This means that, if the system has the DCC in the
iso-singlet channel $\eta_0(x)$ and if $\chi M\simeq 0$ at high enough
$T$, we can interpret this $\eta_0(x)$ as an effective $\theta(x)$ in
a transient state (this reinterpretation exactly corresponds to the
normalization condition for $U$ in Ref.~\cite{Kaiser:2000ck}).  We
note that in the whole argument this is the only place where (partial)
chiral symmetry restoration is required in the hadronic picture of the
CME.  Hence, in the hadronic phase, the DCC of $\eta_0$ is the source
for $\mu_5(x)$.  Its strength and distribution could be in principle
figured out in numerical simulations as in Ref.~\cite{Ikezi:2003kn}.

The anomalous processes such as
$\pi^0\to \gamma\gamma$ and $\gamma \pi^0\to\pi^+ \pi^-$ are described
by the Wess-Zumino-Witten (WZW) part that can be written in a concise
way in the two-flavor case~\cite{Kaiser:2000ck} as
\begin{equation}
 \begin{split}
 & \Lag_{\text{WZW}} = -\frac{\Nc}{32\pi^2}\epsilon^{\mu\nu\rho\sigma}
  \Biggl[ \tr\Bigl\{ U^\dagger\hat{r}_\mu U\hat{l}_\nu
  - \hat{r}_\mu\hat{l}_\nu \\
 & + i\Sigma_\mu (U^\dagger\hat{r}_\nu U \!+\! \hat{l}_\nu) \Bigr\}
  \tr(v_{\rho\sigma}) \!+\! \frac{2}{3} \tr\bigl( \Sigma_\mu \Sigma_\nu
  \Sigma_\rho \bigr)\, \tr(v_\sigma) \Biggr]
 \end{split}
\label{eq:WZW}
\end{equation}
with
$v_{\mu\nu}\equiv\partial_\mu v_\nu-\partial_\nu v_\mu-i[v_\mu,v_\nu]$,
and $\Sigma_\mu \equiv U^\dagger\partial_\mu U$.  A hat symbol
indicates the traceless part, i.e.\
$\hat{r}_\mu \equiv r_\mu - \tfrac{1}{2}\tr(r_\mu)$ and
$\hat{l}_\mu \equiv l_\mu - \tfrac{1}{2}\tr(l_\mu)$.  There is one
more part that has no dynamics of chiral field $U$ and thus is called
the contact part;
\begin{align}
 &\Lag_{\text{P}} = \frac{\Nc}{8\Nf\,\pi^2}\epsilon^{\mu\nu\rho\sigma}
  \Biggl\{ \tr\Bigl[ v_\mu \Bigl( \partial_\nu v_\rho
  -\frac{2i}{3}v_\nu v_\rho \Bigr)\Bigr] \partial_\sigma\theta \\
 & + \tr\bigl( a_\mu D^v_\nu a_\rho \bigr) \Bigl( \frac{4}{3}
  \tr(a_\sigma) + \partial_\sigma \theta \Bigr)
  \!-\!\frac{2}{3\Nf}\tr(a_\mu) \tr(\partial_\nu a_\rho)
  \partial_\sigma \theta \Biggr\}\;, \notag
\end{align}
where $D^v_\mu a_\nu\equiv\partial_\mu a_\nu-i v_\mu a_\nu-ia_\mu v_\nu$.

Now that we have the chiral effective Lagrangian that should encompass
the anomalous processes, it is straightforward to read the current in
the presence of space-time dependent $\theta(x)$ and the
electromagnetic field $A_\mu$.  To this end, in the two-flavor case,
the vector and the axial-vector fields are set to be
\begin{equation}
 v_\mu = eQA_\mu\;, \qquad
 a_\mu = 0
\end{equation}
with the electric-charge matrix,
$Q=\text{diag}(2/3, -1/3)=1/6+\tau_3$.

Let us first simplify $\Lag_{\text{WZW}}$ and $\Lag_{\text{P}}$,
respectively, which are of our central interest.  It should be
mentioned that the quadratic terms of $A_\mu$ vanish due to the
anti-symmetric tensor, $\epsilon^{\mu\nu\rho\sigma}$.  Then, the first
term in Eq.~\eqref{eq:WZW} vanishes and the rest takes the following
form,
\begin{align}
  \Lag_{\text{WZW}} &= -\frac{\Nc \tr(Q)}{32\pi^2}
  \epsilon^{\mu\nu\rho\sigma} 
   \Bigl\{ ie^2 \tr\Big[ \bigl(\Sigma_\mu + \tilde{\Sigma}_\mu\bigr)
   \tau_3 \Bigr] A_\nu\partial_\rho A_\sigma \notag\\
  &\qquad\qquad\qquad\qquad\quad -\frac{2e}{3} \tr\bigl(
   \Sigma_\mu \Sigma_\nu \Sigma_\rho\bigr) A_\sigma \Bigr\} \;,
\end{align}
where we defined $\tilde{\Sigma}_\mu = (\partial_\mu U) U^\dagger$.
Similarly the contact term can become as simple as
\begin{equation}
 \Lag_{\text{P}} = \frac{\Nc e^2 \tr(Q^2)}{8\Nf\,\pi^2}
  \epsilon^{\mu\nu\rho\sigma} A_\mu(\partial_\nu A_\rho)\,
  \partial_\sigma \theta\;.
\label{eq:P}
\end{equation} 


Now, we are ready to confirm that we can reproduce the electric
current corresponding to the CME in the hadronic phase.  We shall next
compute the electric current by taking the differentiation of the
effective action with respect to the gauge field coupled to it, that
is,
\begin{equation}
 j^\mu(x) = \frac{\delta}{\delta A_\mu(x)}
  \int d^4x\,\Lag_{\text{eff}}\;.
\end{equation}
The current from the usual chiral Lagrangian $\Lag_\chi$ at the lowest
order results in
\begin{equation}
 \begin{split}
 j^\mu_\chi &= -i\frac{e f^2_\pi}{4} \tr\bigl[\bigl(\Sigma^\mu -
  \tilde{\Sigma}^\mu\bigr)\tau^3\bigr] \\
 &\simeq e \bigl( \pi^- i\partial^\mu \pi^+
  -\pi^+ i\partial^\mu \pi^- \bigr) + \cdots \;,
 \end{split}
\label{eq:jchi}
\end{equation}
which represents the electric current carried by the flow of charged
pions, $\pi^\pm$, which is clear from the expanded expression.  There
appears no term involving $\partial_\mu\theta$ in this part.  More
non-trivial and interesting is the current associated with the WZW
terms, leading to
\begin{align}
 & j_{\text{WZW}}^\mu = -\frac{\Nc \tr(Q)}{32\pi^2}
  \epsilon^{\mu\nu\rho\sigma} \Bigr\{ 2ie^2 \tr\big[
  (\Sigma_\nu + \tilde{\Sigma}_\nu)\tau_3 \big] \partial_\rho A_\sigma
  \notag\\
 & + e^2 \tr\big[ \partial_\rho(\Sigma_\nu + \tilde{\Sigma}_\nu)
  \tau_3 \big] A_\sigma  - \frac{2e}{3}
  \tr(\Sigma_\nu \Sigma_\rho \Sigma_\sigma) \Bigl\} \;,
\label{eq:jwzw}
\end{align}
The physical meaning of this current will be transparent in the
expanded form using $U\sim 1
+i\boldsymbol{\pi}\cdot \boldsymbol{\tau}/f_\pi+\cdots$.  Then we find
that the first term in Eq.~\eqref{eq:jwzw} is written as,
\begin{equation}
 j_{\text{WZW}}^\mu = \frac{\Nc \tr(Q) e^2}{8\pi^2 f_\pi}\,
  \epsilon^{\mu\nu\rho\sigma}(\partial_\nu \pi^0) F_{\rho\sigma}\;.
\label{eq:jpi0}
\end{equation}
The second term in Eq.~\eqref{eq:jwzw} is vanishing and the last term
represents a topological current purely from the entanglement of all
$\pi^0$ and $\pi^\pm$.  The physics implication of Eq.~\eqref{eq:jpi0}
has been discussed with the $\pi^0$-domain wall~\cite{Son:2007ny} and
the pion profile in the Skyrmion~\cite{Eto:2011id}.  Finally we can
reproduce the CME current from the contact interaction as
\begin{equation}
 j_{\text{P}}^\mu = \frac{\Nc\, e^2\, \tr(Q^2)}{4\Nf\,\pi^2}
  \epsilon^{\mu\nu\rho\sigma} (\partial_\nu A_\rho)\,
  \partial_\sigma \theta \;.
 \label{eq:jp}
\end{equation}
We can rewrite the above expression in a more familiar form using
$\mu_5 = \partial_0\theta/(2\Nf)$ and
$B^i = \epsilon^{ijk} \partial_j A_k$ to reach,
\begin{equation}
 \boldsymbol{j}_{\text{P}} = \frac{\Nc\, e^2\, \tr(Q^2)}{2\pi^2} \mu_5
  \boldsymbol{B} \;.
\label{eq:CME}
\end{equation}
It should be noted that $\epsilon_{0123}=+1$ in our convention.

This derivation of the CME is quite suggestive on its own and worth
several remarks.

First, it is known that the contact term $\Lag_{\text{P}}$ is not
renormalization-group invariant~\cite{Kaiser:2000ck}.  This means that
$\Lag_{\text{P}}$ and thus $j_{\text{P}}$ are scale dependent like the
running coupling constant.  It is often said that $j_{\text{P}}$ is an
exact result from the quantum anomaly, but it may be a little
misleading.  The functional form itself could be protected (though
there is no rigourous proof) but $\bB$ and $\mu_5$ in
Eq.~\eqref{eq:CME} should be renormalized ones.  Indeed it has been
pointed out that interaction vertices in the (axial) vector channels
result in the dielectric correction to
$\bB$~\cite{Gorbar:2009bm,*Fukushima:2010zza}.  The knowledge on the
chiral Lagrangian strongly supports the results of
Ref.~\cite{Gorbar:2009bm}.

Second, to find Eq.~\eqref{eq:CME}, we do not need quark degrees of
freedom explicitly but only hadronic variables.  This is naturally 
so because the idea of the WZW action is to capture the anomalous
effects from the ultraviolet regime in terms of infrared variables.
It is clear from the above derivation, therefore, that the CME occurs
without massless quarks in the quark-gluon plasma.  (See also
Ref.~\cite{Sadofyev:2010pr,*Sadofyev:2010is,*Gao:2012ix,*Son:2012wh}
for another derivations of the CME without referring to quarks
explicitly.)  Then, a conceptual confusion might arise;  what really
flows that contributes to an electric current in the hadronic phase?
One may have thought that it is $\pi^\pm$, but such a current is
rather given by $j_\chi^\mu$, and the CME current $j_{\text{P}}^\mu$
originates from the contact part that is decoupled from $U$.  The same
question is applied to Eq.~\eqref{eq:jpi0} if the system has a $\pi^0$
condensation.

In a sense these currents associated with the $\theta(x)$ or
$\pi^0(x)$ backgrounds are reminiscent of the Josephson current in
superconductivity.  Suppose that we have a $\pi^0$ condensate, then
such a coherent state behaves like a macroscopic wave-function of
$\pi^0$ field.  Then, a microscopic current inside of the
wave-function $\pi^0$ could be a macroscopic current in the whole
system since the wave-function spreads over the whole system.  In the
case of the CME, $\theta(x)$ or $\eta_0(x)$ plays the same role as
$\pi^0(x)$.  In this way, strictly speaking, it is a high-momentum
component of quarks and anti-quarks in the wave-function of $\pi^0$ or
$\eta_0$ that really flow to make a current, though these quarks do
not have to get deconfined.


This sort of confusing interpretation of the CME current arises from
the assumption that $\theta(x)$ and $\bB(x)$ are spatially
homogeneous.  Once this assumption is relaxed, as we discuss in what
follows, an interesting new possibility opens, which may be more
relevant to experiments.

\begin{figure}
 \includegraphics[width=0.6\columnwidth]{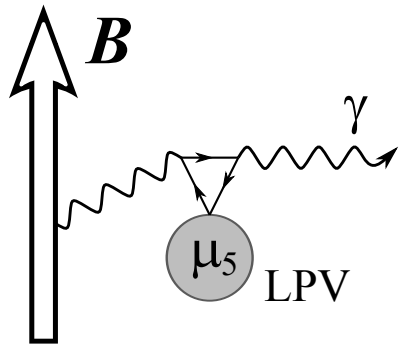}
 \caption{Schematic figure for the single photon production as a
   consequence of the axial anomaly and the external magnetic field.
   The angular distribution of the emitted photons is proportional to
   $(q_z^2+q_x^2)/(q_x^2+q_y^2+q_z^3)$ where $q_y$ is in the direction
   parallel to $\bB$ and $q_z$ and $q_x$ are perpendicular to $\bB$.}
 \label{fig:photon}
\end{figure}

From now on, let us revisit Eq.~\eqref{eq:P} from a different point of
view.  If we literally interpret Eq.~\eqref{eq:P} as usual in the
quantum field theory, it should describe a vertex of the processes
involving two photons and the $\theta$ field such as $\theta \to
\gamma\gamma$ and $\theta + B \to \gamma$ in the magnetic field.  The
latter process can be viewed as the reverse of the Primakoff effect
involving the $\theta(x)$ background instead of neutral mesons.  It is
a very intriguing question how much photon can be produced from this
reverse Primakoff effect.  For this purpose we shall decompose the
vector potential into the background $\bar{A}_\mu$ (corresponding to
$B$) and the fluctuation $\calA_\mu$ (corresponding to photon).  Then,
Eq.~\eqref{eq:P} turns into
\begin{equation}
 \Lag_{\text{P}} = \frac{\Nc\, e^2\, \tr(Q^2)}{8\Nf\,\pi^2}\,
  \epsilon^{\mu\nu\rho\sigma} \bigl[
  \calA_\mu(\partial_\nu \calA_\rho) + \calA_\mu \bar{F}_{\nu\rho}
  \bigr] \partial_\sigma\theta \;,
\label{eq:vertex}
\end{equation}
where the first term represents the two-photon production process
$\theta\to \gamma\gamma$ similar to $\pi^0\to \gamma\gamma$, and the
second represents the reverse Primakoff effect
($\theta + B \to \gamma$) involving the background field strength
$\bar{F}_{\mu\nu}=\partial_\mu\bar{A}_\nu-\partial_\nu\bar{A}_\mu$.
Here we are interested only in the situation that the background field
is so strong that we can neglect the contribution from the first
term.

Even when $|eB|\sim \LQCD$ in the heavy-ion collision, we can still
utilize the perturbative expansion in terms of the electromagnetic
coupling constant.  In the leading order, from the LSZ reduction
formula, the $S$-matrix element for the single-photon production with
the momentum $q=(|\bq|,\bq)$ and the polarization
$\varepsilon^{(i)}(\bq)$ is deduced immediately from the
vertex~\eqref{eq:vertex},
\begin{equation}
 \begin{split}
 &i\calM(i;\bq) = \langle \varepsilon^{(i)}(\bq)|\Omega\rangle
   = i \frac{\Nc\, e^2\, \tr(Q^2)}
  {8\Nf\,\pi^2\sqrt{(2\pi)^3 2q_0}} \\
 &\qquad\times
  \epsilon^{\mu\nu\rho\sigma} \varepsilon^{(i)\mu}(\bq)
  \int d^4 x\,e^{-iq\cdot x} \bar{F}_{\nu\rho}(x)\,
  \partial_\sigma \theta(x) \;,
 \end{split}
\end{equation}
where $q_0 = |\boldsymbol{q}|$.  This expression becomes very simple
when the background field has only the magnetic field in the $y$
direction, i.e.\ $B=\bar{F}_{zx}$ and the rest is just vanishing.
Thus, we have,
\begin{equation}
 \begin{split}
  &\epsilon^{\mu\nu\rho\sigma} \varepsilon^{(i)\mu}(\bq)
  \int d^4 x\,e^{-iq\cdot x} \bar{F}_{\nu\rho}(x)\,
  \partial_\sigma \theta(x) \\
  &\qquad\qquad = -2\varepsilon^{(i) y}(\bq)\int d^4x\, e^{-iq\cdot x}
  B(x)\, \partial_0 \theta(x)\;,
 \end{split}
\end{equation}
and replacing $\partial_0\theta$ by the chiral chemical potential
$\mu_5$ by $\mu_5=\partial_0\theta/(2\Nf)$ and using
$\sum_i \varepsilon^{(i)j}(\bq)\,\varepsilon^{(i)k}(\bq)
=\delta^{jk}-q^j q^k/\bq^2$ with $\bq^2=q_x^2+q_y^2+q_z^2$, we can
finally arrive at
\begin{align}
 &q_0 \frac{dN_\gamma}{d^3q} = q_0 \sum_i |\calM(i;\bq)|^2 \notag\\
 &= \frac{1-(q_y)^2/\bq^2}{2 (2\pi)^3}
  \biggl( \frac{\Nc\, e^2\, \tr(Q^2)}{2\pi^2}\int d^4x\, e^{-iq\cdot x}
  B(x)\mu_5(x)\biggr)^2 \notag\\
 &= \frac{q_z^2+q_x^2}{2 (2\pi)^3 \bq^2}
  \cdot \frac{25\,\alpha_e\, \zeta(\bq)}{9\pi^3} \;,
\end{align}
where we used $\Nc=3$ and $\tr(Q^2)=5/9$ for the two-flavor case in
the last line and $\alpha_e\equiv e^2/(4\pi)\simeq 1/137$ is the fine
structure constant.  In the above we defined,
\begin{equation}
 \zeta(\bq)\equiv \biggl|\int d^4x\,e^{-iq\cdot x} eB(x)\,
  \mu_5(x)\biggr|^2 \;.
\end{equation}

It is quite interesting to see that the final expression is
proportional to the momenta $q_z^2+q_x^2$ which are perpendicular to
the $\bB$ direction.  This could be another source for the elliptic
flow $v_2$ of the direct photon in a similar mechanism as pointed out
in Ref.~\cite{Basar:2012bp}.

Because there is no reliable model to predict $\mu_5(x)$, it is
difficult to calculate $\zeta(\bq)$ as a function of the momentum.
For a first attempt, therefore, let us make a qualitative order
estimate.  The strength of the magnetic field is as large as $\LQCD^2$
or even bigger at initial time.  A natural scale for $\mu_5$ is also
given by $\LQCD$, or if the origin of the LPV is the color flux-tube
structure in the Glasma~\cite{Kharzeev:2001ev}, the typical scale is
the saturation momentum $Q_s\sim 2\;\text{GeV}$ for the RHIC energy.
The space-time integration picks up the volume factor $\sim \tau_0^2
A_\perp$ with $\tau_0$ being the life time of the magnetic field, i.e.\
$\tau_0\simeq 0.01\sim 0.1\;\text{fm/c}$, and $A_\perp$ the transverse
area $\sim 150\;\text{fm}^2$ for the Au-Au collision.  Then, $\zeta
\simeq 0.1\sim 10^{3}$, where the smallest estimate for
$\tau_0=0.01\;\text{fm/c}$ and $\mu_5\sim\LQCD$ and the largest one
for $\tau_0=0.1\;\text{fm/c}$ and $\mu_5\sim Q_s$.  Then, the photon
yield is expected to be
$q_0 (dN_\gamma/d^3q) \simeq (10^{-7} \sim 10^{-3})\text{GeV}^{-2}$.
This is a rather conservative estimate, for the magnetic field may
live longer with backreactions and is of detectable level of the
photon yield as compared to the conventional photon production from
the thermal medium~\cite{Alam:2000bu,*Srivastava:2001hz,%
*Rasanen:2002qe,*Gelis:2004ep,*Turbide:2003si,*Turbide:2007mi}.

One may think that not only the polarization but also $\zeta(\bq)$ has
strong asymmetry because of the presence of $\bB$.  The typical domain
size of the LPV should be, however, much smaller than the impact
factor $b\sim$ a few fm at least, and thus
the asymmetry effect turns out only negligible.  In reality, depending
on the spatial position, there are not only $B_y$, but $B_x$ and $B_z$
and also the electric fields $E_x$, $E_y$, and $E_z$.  We are now
performing full numerical calculations including all those fields and
the LPV based on the Glasma flux-tube picture.  Since such model
buildings postulate lots of arguments on assumptions and
justifications, we will leave them to a separate publication under
preparation.

In summary, we have formulated the CME in terms of the chiral
Lagrangian with the WZW terms, which provides us with the physics
picture to understand the CME in the hadronic phase.  We derived the
current of the CME correctly from the contact term that is not RG
invariant.  We established how the CME could be realized through
$\eta_0(x)$ as a result of the DCC in the iso-singlet channel.  Then,
the key observation in view of the chiral Lagrangian is that the
vertex responsible for the CME also describes the single photon
production for space-time inhomogeneous $\theta(x)$ and $\bB(x)$.  We
have given the expression for the photon yield to find that its
angular distribution is asymmetric with the direction perpendicular to
$\bB$ more preferred.  We made a qualitative estimate for the yield
and found it comparable to the thermal photon contribution.  Unlike
the thermal photon the $p_t$ distribution should reflect the domain
size of the LPV.\ \ Electromagnetic probes as a signature for the LPV
(see Ref.~\cite{Andrianov:2010ah,*Andrianov:2012hq} for the dilepton
production) deserve further investigations and we believe that this
work would shed light on future developments in this direction.

\begin{acknowledgments}
We thank Dima~Kharzeev, Naoki~Yamamoto, and Qun~Wang for useful
discussions.  K.~F.\ is grateful for discussions in a workshop ``Heavy
Ion Pub'' at Hiroshima University, which inspired him.  K.~F.\ is
supported by Grant-in-Aid for Young Scientists B (24740169).
\end{acknowledgments}


\bibliography{photon}

\end{document}